\def\figuresizeLarge{6.5in}

\def\expectationOperator[#1][#2]{\mathbb{E}\left[#1\right]}
\def\uniformDistribution[#1][#2]{{\mathcal{U}_{[#1,#2]}}}
\def\traceOperator[#1]{{\mathrm{tr}}\{#1\}}
\def\identityMatrix[#1]{\textbf{\mathrm{I}}_{#1}}
\def\zeroVector[#1]{\textbf{\mathrm{0}}_{#1}}

\def\clamp[#1][#2]{\text{clamp}_{#2}\left(#1\right)}
\def\signNormal[#1]{\text{sign}\left(#1\right)}
\def\diagOperation[#1]{\text{diag}\left\{#1\right\}}
\def\clamp[#1][#2]{\text{clamp}_{#2}\left(#1\right)}

\def\probability[#1]{\mathrm{Pr}\left({#1}\right)}
\def\complexGaussian[#1][#2]{\mathcal{CN}({#1,#2})}
\def\gaussian[#1][#2]{\mathcal{N}({#1,#2})}

\def\normalPDF[#1]{\phi\left(#1\right)}
\def\normalCDF[#1]{\Phi\left(#1\right)}
\def\CDF[#1][#2][#3]{F_{#1}^{#2}\left({#3}\right)}
\def\PDF[#1][#2][#3]{f_{#1}^{#2}\left({#3}\right)}
\def\indicatorFunction[#1]{\mathbb{I}\left[{#1}\right]}
\def\functionConflict[#1][#2]{\mathcal{M}_{#1}{(#2)}}
\def\indexNode{n}
\def\indexNodeOther{m}
\def\indexNodeAnother{l}

\def\numberOfNodesInside{N_{\rm sen}}
\def\numberOfNodesReference{N_{\rm ref}}
\def\numberOfNodesTotal{N_{\rm node}}
\def\nodeID[#1]{a_{#1}}
\def\nodePosition[#1]{{\rm \bf p}_{#1}}
\def\nodePositionEstimated[#1]{{\rm \bf \tilde{p}}_{#1}}
\def\nodePositionX[#1]{{x}_{#1}}
\def\nodePositionY[#1]{{y}_{#1}}
\def\nodePositionZ[#1]{{z}_{#1}}
\def\nodeSlot[#1][#2]{s_{#1}^{(#2)}}
\def\nodeSlotOpt[#1][#2]{\hat{s}_{#1}^{(#2)}}
\def\nodeTransmitPower[#1]{P_{#1}}
\def\nodeDiscovered[#1][#2]{d_{#1}^{(#2)}}
\def\nodeHoppingNumber[#1]{h_{#1}^{(\globalTime)}}
\def\hoppingNumberMax{h_{\rm N/A}}

\def\distance[#1][#2]{d_{#1,#2}}
\def\neighborSetR[#1]{\mathcal{\acute{R}}_{#1}^{(\globalTime)}}
\def\neighborSet[#1]{\mathcal{\acute{N}}_{#1}^{(\globalTime)}}
\def\neighborSetRX[#1][#2]{\mathcal{R}_{#1}^{(#2)}}
\def\neighborSetTX[#1][#2]{\mathcal{N}_{#1}^{(#2)}}
\def\numberOfSlots{N_{\rm slot}}

\def\durationSlot{T_{\rm slot}}
\def\durationComm{T_{\rm comm}}
\def\durationProcessing{T_{\rm proc}}

\def\durationPacket{T_{\rm init}}
\def\durationBeacon{T_{\rm beacon}}
\def\durationRemaining[#1]{T_{{\rm state},#1}^{(\globalTime)}}
\def\durationDiff{T_{\rm r}}
\def\globalTime{t}

\def\probabilityInitiator{p}
\def\arandomVariableForInit[#1]{r_{#1}^{(\globalTime)}}
\def\accessible[#1][#2]{c_{#1,#2}^{(\globalTime)}}

\def\nodeState[#1]{S_{#1}^{(\globalTime)}}

\def\stateProcessing{{\tt{P}}}
\def\stateResOne{{\tt{R}_1}}
\def\stateResTwo{{\tt{R}_2}}
\def\stateInit{{\tt{I}}}

\def\nForget{N_{\rm max}}
\def\aDuration{T}

\documentclass[conference]{IEEEtran}
\IEEEoverridecommandlockouts
\usepackage{multicol}
\usepackage{lipsum}
\usepackage{mathtools}
\usepackage{amsthm}
\usepackage{acronym}
\usepackage{stfloats}
\usepackage{amsfonts}
\usepackage{acronym}
\usepackage[noadjust]{cite}
\usepackage{multirow}
\usepackage{bm}
\usepackage{amsmath,amssymb}
\usepackage{graphicx}
\usepackage{epstopdf}
\usepackage[table,xcdraw]{xcolor}
\usepackage[caption=false,font=footnotesize]{subfig}
\usepackage[geometry]{ifsym}
\usepackage{array}
\usepackage[utf8]{inputenc}
\usepackage[T1]{fontenc}
\usepackage{pifont}
\usepackage{tabularray}
\usepackage{lipsum}
\newcommand\mydots{\hbox to 1em{.\hss.\hss.}}

\makeatletter
\newif\ifAC@uppercase@first%
\def\Aclp#1{\AC@uppercase@firsttrue\aclp{#1}\AC@uppercase@firstfalse}%
\def\AC@aclp#1{%
	\ifcsname fn@#1@PL\endcsname%
	\ifAC@uppercase@first%
	\expandafter\expandafter\expandafter\MakeUppercase\csname fn@#1@PL\endcsname%
	\else%
	\csname fn@#1@PL\endcsname%
	\fi%
	\else%
	\AC@acl{#1}s%
	\fi%
}%
\def\Acp#1{\AC@uppercase@firsttrue\acp{#1}\AC@uppercase@firstfalse}%
\def\AC@acp#1{%
	\ifcsname fn@#1@PL\endcsname%
	\ifAC@uppercase@first%
	\expandafter\expandafter\expandafter\MakeUppercase\csname fn@#1@PL\endcsname%
	\else%
	\csname fn@#1@PL\endcsname%
	\fi%
	\else%
	\AC@ac{#1}s%
	\fi%
}%
\def\Acfp#1{\AC@uppercase@firsttrue\acfp{#1}\AC@uppercase@firstfalse}%
\def\AC@acfp#1{%
	\ifcsname fn@#1@PL\endcsname%
	\ifAC@uppercase@first%
	\expandafter\expandafter\expandafter\MakeUppercase\csname fn@#1@PL\endcsname%
	\else%
	\csname fn@#1@PL\endcsname%
	\fi%
	\else%
	\AC@acf{#1}s%
	\fi%
}%
\def\Acsp#1{\AC@uppercase@firsttrue\acsp{#1}\AC@uppercase@firstfalse}%
\def\AC@acsp#1{%
	\ifcsname fn@#1@PL\endcsname%
	\ifAC@uppercase@first%
	\expandafter\expandafter\expandafter\MakeUppercase\csname fn@#1@PL\endcsname%
	\else%
	\csname fn@#1@PL\endcsname%
	\fi%
	\else%
	\AC@acs{#1}s%
	\fi%
}%
\edef\AC@uppercase@write{\string\ifAC@uppercase@first\string\expandafter\string\MakeUppercase\string\fi\space}%
\def\AC@acrodef#1[#2]#3{%
	\@bsphack%
	\protected@write\@auxout{}{%
		\string\newacro{#1}[#2]{\AC@uppercase@write #3}%
	}\@esphack%
}%
\def\Acl#1{\AC@uppercase@firsttrue\acl{#1}\AC@uppercase@firstfalse}
\def\Acf#1{\AC@uppercase@firsttrue\acf{#1}\AC@uppercase@firstfalse}
\def\Ac#1{\AC@uppercase@firsttrue\ac{#1}\AC@uppercase@firstfalse}
\def\Acs#1{\AC@uppercase@firsttrue\acs{#1}\AC@uppercase@firstfalse}

\acrodef{WSN}{wireless sensor network}
\acrodef{ID}{identity}
\acrodef{USRP}{universal software radio peripheral}
\acrodef{SN}{sensor node}
\acrodef{FC}{fusion center}
\acrodef{MAC}{multiple-access channel}
\acrodef{FL}{federated learning}
\acrodef{ED}{edge device}
\acrodef{CS}{compressed sensing}
\acrodef{ES}[BS]{base station}
\acrodef{DCN}{data center network}
\acrodef{RIS}{reconfigurable intelligent surfaces}
\acrodef{IMC}{in-memory computing}
\acrodef{FPGA}{field-programmable gate array}
\acrodef{SDR}{software-defined radio}
\acrodef{GPS}{Global Positioning System}
\acrodef{PS}{processing system}
\acrodef{SS}{soft synchronization}
\acrodef{IQ}{in-phase/quadrature}
\acrodef{IP}{intellectual property}
\acrodef{DMA}{direct-memory access}
\acrodef{RAM}{random access memory}
\acrodef{CC}{companion computer}
\acrodef{FEE}{function estimation error}
\acrodef{MSK}{minimum-shift keying}
\acrodef{TDMA}{time-domain multiple access}
\acrodef{PLNC}{physical-layer network coding}
\acrodef{UAV}{unmanned aerial vehicle}
\acrodef{LoRa}{Long-Range}
\acrodef{DC}{direct-current}
\acrodef{DAC}{digital-to-analog converter}
\acrodef{ADC}{anlog-to-digital converter}
\acrodef{CS}{complementary sequence}
\acrodef{GCP}{Golay complementary pair}
\acrodef{ANF}{algebraic normal form}
\acrodef{AACF}{aperiodic auto-correlation function}
\acrodef{RM}{Reed-Muller}
\acrodef{LDI}{localization-done indicator}
\acrodef{PUCCH}{physical uplink control channel}
\acrodef{GUI}{graphical user interface}
\acrodef{OBO}{output-power back-off}
\acrodef{ACLR}{adjacent-channel-leakage ratio}

\acrodef{LDPC}{low-density parity check}

\acrodef{PDF}{probability density function}
\acrodef{CDF}{cummulative distribution function}

\acrodef{LoRaWAN}{Long Range Wide Area Network}
\acrodef{TBMA}{type-based multiple access}
\acrodef{TDD}{time-domain duplexing}

\acrodef{MSFE}{mean-squared function error}
\acrodef{FEE}{function-estimation error}
\acrodef{CER}{computation error rate}
\acrodef{BCER}{block-computation error rate}
\acrodef{CFO}{carrier frequency offset}
\acrodef{TO}{time offset}
\acrodef{PO}{phase offset}
\acrodef{RSSI}{received signal strength  information}
\acrodef{CSMA/CA}{carrier-sense multiple access with collision avoidance}
\acrodef{RTToF}{round trip time-of-flight}

\acrodef{STLC}{space-time line code}
\acrodef{CCI}{co-channel interference}
\acrodef{CSIT}[CSIT]{\ac{CSI} at the transmitter}
\acrodef{CSIR}[CSIR]{\ac{CSI} at the receiver}
\acrodef{MIMO}{multiple-input-multiple-output}
\acrodef{PC}{phase correction}
\acrodef{ZF}{zero-forcing}
\acrodef{ANOVA}{analysis of variance}

\acrodef{PCA}{principal component analysis}
\acrodef{TIG}{Technical Interest Group}

\acrodef{FSK}{frequency-shift keying}
\acrodef{PPM}{pulse-position modulation}
\acrodef{PAM}{pulse-amplitude modulation}

\acrodef{MRC}{maximum-ratio combining}
\acrodef{HP}{hard-coded participation}
\acrodef{HPA}{hard-coded participation with absentees}
\acrodef{SP}{soft-coded participation}
\acrodef{FSK-MV}{\ac{FSK}-based \ac{MV}}
\acrodef{RF}{radio-frequency}
\acrodef{MF}{matched filter}
\acrodef{PPM}{pulse-position modulation}
\acrodef{CSK}{chirp-shift keying}
\acrodef{PPM-MV}[PPM-MV]{\ac{PPM}-based \ac{MV}}
\acrodef{DFT-s-OFDM}{\ac{DFT}-spread \ac{OFDM}}
\acrodef{SC}{single-carrier}
\acrodef{SGD}{stochastic gradient descent}
\acrodef{signSGD}{sign stochastic gradient descent}

\acrodef{SL}{split learning}
\acrodef{SNR}{signal-to-noise ratio}
\acrodef{RMSE}{root-mean-squared error}
\acrodef{OFDM}{orthogonal frequency division multiplexing}
\acrodef{DFT}{discrete Fourier transform}
\acrodef{PSK}{phase-shift keying}
\acrodef{QAM}{quadrature amplitude modulation}
\acrodef{QPSK}{quadrature phase-shift keying}
\acrodef{PMEPR}{peak-to-mean envelope power ratio}
\acrodef{BER}{bit-error ratio}
\acrodef{SNR}{signal-to-noise ratio}
\acrodef{PSD}{power spectral density}
\acrodef{SE}{spectral efficiency}
\acrodef{CP}{cyclic prefix}
\acrodef{AWGN}{additive white Gaussian noise}
\acrodef{CFR}{channel frequency response}
\acrodef{CIR}{channel impulse response}
\acrodef{MMSE}{minimum mean-squared error}
\acrodef{LMMSE}{linear minimum mean-squared error}
\acrodef{BPSK}{binary phase shift keying}
\acrodef{BPSK}{quadrature phase shift keying}
\acrodef{BLER}{block-error rate}
\acrodef{ML}{maximum likelihood}
\acrodef{PHY}{physical layer}
\acrodef{PA}{power amplifier}
\acrodef{IDFT}{inverse DFT}
\acrodef{DoF}{degrees-of-freedom}
\acrodef{IoT}{Internet-of-Things}
\acrodef{FDE}{frequency-domain equalization}
\acrodef{RF}{radio-frequency}
\acrodef{IM}{index modulation}
\acrodef{MF}{matched filter}
\acrodef{PPM}{pulse-position modulation}

\acrodef{MSE}{mean-squared error}
\acrodef{MRT}{maximum-ratio transmission}
\acrodef{ERC}{equal-ratio combining}
\acrodef{BAA}{broadband analog aggregation}
\acrodef{OBDA}{one-bit broadband digital aggregation}
\acrodef{FEEL}{federated edge learning}
\acrodef{FL}{federated learning}
\acrodef{UL}{uplink}
\acrodef{DL}{downlink}
\acrodef{OAC}{over-the-air computation}
\acrodef{TCI}{truncated-channel inversion}
\acrodef{MV}{majority vote}
\acrodef{CNN}{convolution neural network}
\acrodef{ReLU}{rectified-linear unit}
\acrodef{CSI}{channel state information}
\acrodef{PAPR}{peak-to-average power ratio}
\acrodef{SC}{single-carrier}
\acrodef{iid}[IID]{independent and identically distributed}
\acrodef{RMS}{root-mean-square}
\acrodef{4G}{Fourth Generation}
\acrodef{5G}{Fifth Generation}
\acrodef{NR}{New Radio}
\acrodef{LTE}{Long-Term Evolution}
\acrodef{DFT-s-OFDM}{\ac{DFT}-spread \ac{OFDM}}
\acrodef{OFDMA}{orthogonal frequency division multiple access}
\acrodef{HARQ}{hybrid automatic repeat request}
\acrodef{D2D}{Device-to-Device}
\acrodef{NOMA}{non-orthogonal multiple access}
\acrodef{OMA}{orthogonal multiple access}

\acrodef{IMT}{International Mobile Telecommunications}
\acrodef{ITU}{International Telecommunication Union}
\acrodef{NLoS}{non-line-of-sight}
\acrodef{PAR}{periodically alternating roles}

\begin{document}

\title{
	{A Self-Healing Mesh Network without Global-Time Synchronization}
	\\
\thanks{This research is supported by TUBITAK BIDEB 2221 Program 2023.}
\author{
	\IEEEauthorblockN{
		Alphan \c{S}ahin\IEEEauthorrefmark{1} and H\"{u}seyin Arslan\IEEEauthorrefmark{2}
	} \IEEEauthorblockA{\IEEEauthorrefmark{1}University of South Carolina, Columbia and \IEEEauthorrefmark{2}Istanbul Medipol University, Istanbul\\
		E-mails: 
		asahin@mailbox.sc.edu,  
		huseyinarslan@medipol.edu.tr
	}
} 
}

\maketitle


\maketitle

\begin{abstract}
In this paper, we propose a slot-based protocol that does not rely on global-time synchronization to achieve a self-healing mesh network. With the proposed protocol, each node  synchronizes with its neighbors locally by adjusting its time to transmit based on the reception instant of a decoded beacon signal. Also, it determines its slots without any coordinator to avoid collisions. Finally, to communicate the messages over the mesh network, it identifies the forwarding nodes on the shortest path without knowing the entire communication graph. We show that the proposed protocol can effectively resolve  collisions over time while enabling nodes to synchronize with each other in a distributed manner. We numerically analyze the performance of the proposed protocol for different configurations under a realistic channel model considering asymmetrical links. We also implement the proposed method in practice with \ac{LoRa} devices. We demonstrate that the nodes adapt themselves to changes in the network and deliver a message from a sensing node to a reference node via multi-hop routing.

\end{abstract}
\acresetall
\section{Introduction}

\ac{LoRa} devices are often used via \ac{LoRaWAN} standard relying on a star  topology, where the nodes communicate with the gateways connected to the Internet \cite{Jouhari_2023}. 
The gateways forward the data acquired from nodes to the central servers, and users can interact with the central servers by accessing the data. Although this approach is useful for a wide range of applications, the coverage area of a LoRaWAN network is limited to the coverage area of the gateways. Also, the coverage range of a typical LoRa node can degrade in \ac{NLoS} conditions since  the narrowband LoRa signals can be affected severely by the fading in multi-path channels. Hence, a star topology may not address the communication scenarios where the signals are obstructed. To address the coverage problem, an alternative solution is a mesh topology \cite{Centelles_2021,Cotrim_2020}. With mesh, the communication is achieved by routing the packets over multiple nodes. However, building a large-scale reliable mesh network with low-cost LoRa devices in a distributed manner is not a trivial  task because of the collision problems, lack of time-synchronization among nodes, asymmetrical links due to the non-identical transmit powers or hardware impairments, time-variant communication channels, low-end microprocessors, and intermittent nodes. In this study, we address these issues with a new protocol compatible with LoRa devices. 

In the literature,  LoRa-based mesh networks have been receiving increasing attention. For instance, in \cite{Heon_2019}, it is proposed to modify \ac{LoRaWAN} with an event-driven approach to support a mesh network. In \cite{Zhu_2019}, parallel transmissions with multiple spreading factors are exploited and the nodes are assigned to several subnets to improve the capacity of the mesh network. 
In \cite{Riccardo_2021}, a peer-to-peer mesh network without using LoRaWAN is proposed, where the clocks of the nodes are assumed to be synchronized.
In \cite{Christian_2019}, to achieve a synchronous LoRa mesh network, the authors utilize the \ac{GPS} or DCF77 long-wave time signaling.
Similarly, in \cite{Manzoni_2023}, nodes acquire real-time clock information from auxiliary signals and broadcast the acquired information for synchronization.
A useful library that allows LoRa nodes to keep the routing tables updated based on routing messages among the nodes  is developed  in \cite{Miquel_2022}  for mesh networks. 
In \cite{Dixin_2023}, the CottonCandy protocol, where the nodes organize themselves in a spanning-tree topology in a distributed fashion, is proposed. In this protocol, the root is a gateway device and all nodes, including the root, start their duty cycles at
about the same time. The nodes are assumed to be roughly synchronized within the range of seconds. Meshtastic is one of the successful open-source  LoRa mesh projects. It is shown that it can support more than 200~km of communication range  \cite{meshtastic_2023}. However, this project adopts a plain flooding approach, and  the underlying protocols are not particularly optimized for communications in NLoS. 
We refer the readers to \cite{Centelles_2021} and the references therein for further discussions on LoRa mesh networks.

In this study, to achieve a self-healing mesh network where the nodes adapt themselves to the changes in the network, we propose a slot-based protocol that locally synchronizes the nodes with their neighbors by exploiting the reception instants of the beacons while  allowing nodes to choose their slots to mitigate the collisions. We also introduce a technique for a node to obtain its hopping distance to a reference node (e.g., a gateway) to find the shortest routing without knowing the communication graph. With comprehensive simulations and a proof-of-concept demonstration with LoRa devices, we show the proposed protocol resolves the collision events over time, synchronizes the nodes without \ac{GPS} signals, and coordinates a dynamic network in a distributed manner.

\section{Problem Statement}
Consider a scenario where $\numberOfNodesInside$ sensing nodes spread in an area. Also, suppose that $\numberOfNodesReference$ reference nodes are deployed in the same region. We assume that all nodes use the same carrier frequency and can either transmit {\em or} receive  signals at a given time. The ultimate goal of the network formed by all nodes is to transfer a  message initiated by a sensing node to one of the reference nodes or vice versa, reliably.

\subsection{Finding and tracking neighbors}
In practice, the link between any two nodes can be severely affected by the wireless channel. Also, the transmit power of the nodes may not be identical. Hence, without loss of generality, the network can be abstracted as a directed dynamic graph changing over time, where the vertices and the arcs correspond to the nodes and the directed communication links, respectively. To express the graph, we let the positive integer $\nodeID[\indexNode]$ to denote the $\indexNode$th node \ac{ID} for $\nodeID[\indexNode]\neq\nodeID[\indexNodeOther]$, $\forall\indexNode,\indexNodeOther\in\{1,2,\mydots,\numberOfNodesTotal\}$ and $\numberOfNodesTotal\triangleq\numberOfNodesInside+\numberOfNodesReference$. 
 Let $\accessible[\indexNode][\indexNodeOther]$ show if the $\indexNode$th node can decode the $\indexNodeOther$th node's packet or not at time $\globalTime$.
If the $\indexNodeOther$th node's signals can be decoded by the $\indexNode$th node, $\accessible[\indexNode][\indexNodeOther]$ is $1$. Otherwise, $\accessible[\indexNode][\indexNodeOther]$ is set to $0$. 
We define the set of nodes that can reach the $\indexNode$th node and the set of nodes that can communicate bi-directionally with the $\indexNode$th node as   $\neighborSetR[\indexNode]\triangleq\{\nodeID[\indexNodeOther]|\accessible[\indexNode][\indexNodeOther]=1,\forall\indexNodeOther\in\{1,2,\mydots,\numberOfNodesTotal\}\}$ and  $\neighborSet[\indexNode]\triangleq\{\nodeID[\indexNodeOther]|\accessible[\indexNode][\indexNodeOther]=\accessible[\indexNodeOther][\indexNode]=1,\forall\indexNodeOther\in\{1,2,\mydots,\numberOfNodesTotal\}\}$, respectively. Hence,  $\neighborSet[\indexNode]$ is a subset of $\neighborSetR[\indexNode]$ in general. 

The first challenge that needs to be addressed is that $\neighborSet[\indexNode]$ and  $\neighborSetR[\indexNode]$ are not available to the $\indexNode$th node. Hence, the $\indexNode$th node needs to learn and actively track the sets $\neighborSetR[\indexNode]$  and $\neighborSet[\indexNode]$  to achieve a self-healing network. To this end, let $\neighborSetRX[\indexNode][\globalTime]$ and $\neighborSetTX[\indexNode][\globalTime]$ denote the set of discovered nodes that can be heard by the $\indexNode$th  node  at time $\globalTime$ and the set of discovered nodes that can communicate with the $\indexNode$th node at time $\globalTime$, respectively. Both $\neighborSetRX[\indexNode][\globalTime]$ and $\neighborSetTX[\indexNode][\globalTime]$  are empty sets for $\globalTime=0$ and need to be obtained by a protocol, as close as possible to $\neighborSet[\indexNode]$ and  $\neighborSetR[\indexNode]$. 

\subsection{Synchronizing nodes and mitigating collisions}
\begin{figure}[t]
	\centering
	{\includegraphics[width=3.0in]{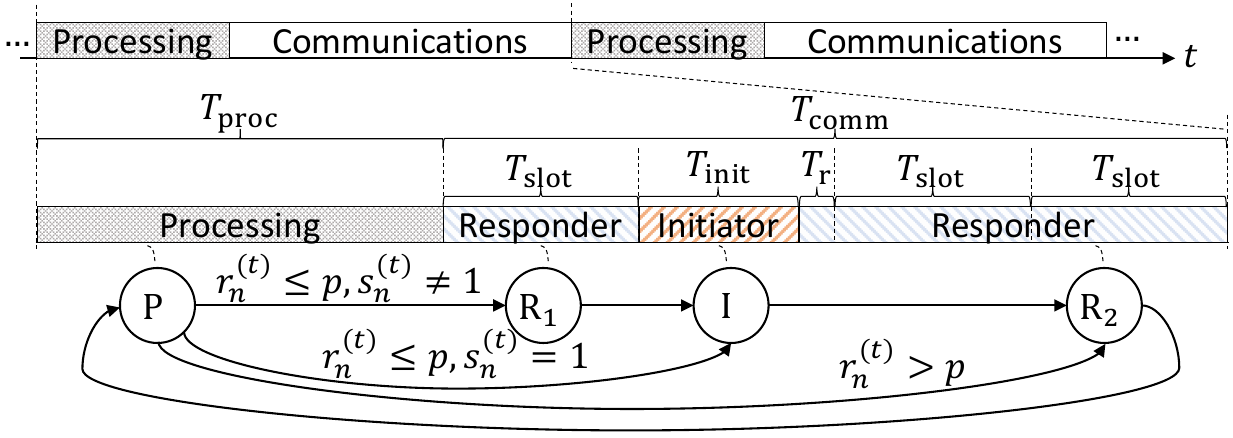}
	} 
	\caption{PAR for $\numberOfSlots=4$ and the corresponding states. In this example, the node uses $\nodeSlot[\indexNode][\globalTime]=2$ when it is an initiator  and changes its state when $\durationRemaining[\indexNode]$ is zero in addition to the conditions given in the state diagram.}
	\label{fig:periodicRoles}
\end{figure}

Due to the absence of global-time synchronization among nodes, a transmitted packet may not be received by the intended receiver as the node may not be listening to the channel at that particular time. This issue can be more severe with typical LoRa nodes as they are often equipped with low-end microprocessors to increase battery life. Hence, the microprocessor  may  need to allocate some time to handle other functionalities, such as a computation task for sensing, causing blind periods. To address this issue, we consider {\em \ac{PAR}}, where we allocate fixed durations for processing, i.e., $\durationProcessing$, and communications, i.e., $\durationComm$, and alternate them. We further divide the communication duration into $\numberOfSlots$ slots, where the duration of each slot is $\durationSlot=\durationComm/\numberOfSlots$ as can be seen in \figurename~\ref{fig:periodicRoles}. Let $\nodeSlot[\indexNode][\globalTime]\in\{1,\mydots,\numberOfSlots\}$  denote the slot index chosen by the $\indexNode$th node at time $\globalTime$. We define three roles for all nodes: 1)~Processing: The node cannot transmit or receive as the microprocessor is busy with some other processing tasks. 2)~Responder: The node actively listens to the communication channel to receive a packet. 3)~Initiator: The node initiates a task by transmitting a packet.  We denote the initiator duration as $\durationPacket$ for $\durationPacket\le\durationSlot$. For $\durationDiff\triangleq\durationSlot-\durationPacket>0$, the node behaves as a responder for the remaining duration within the slot.

In this study, we assume that each node chooses its slots randomly after it wakes up. Also,
 the $\indexNode$th node is granted to be an initiator only within the slot $\nodeSlot[\indexNode][\globalTime]$ with the probability of $\probabilityInitiator$. To express  this probabilistic behavior, let $\arandomVariableForInit[\indexNode]$ be a uniform random variable between $0$ and $1$, drawn during the processing duration. If $\arandomVariableForInit[\indexNode]>\probabilityInitiator$ holds, the node behaves as a responder on the corresponding communication duration. Otherwise, it is granted to be an initiator.  
 To show the transitions between the aforementioned roles, we denote the $\indexNode$th node state as $\nodeState[\indexNode]\in\{\stateProcessing,\stateResOne,\stateInit,\stateResTwo\}$, where $\stateProcessing$, $\stateResOne$, $\stateInit$, and $\stateResTwo$ symbolize the processing, the responder role before being an initiator, the initiator role, and the responder role after being an initiator, respectively. Let $\durationRemaining[\indexNode]$ denote the remaining time for the $\indexNode$th node at time $\globalTime$ to switch its behavior for the next role. When $\durationRemaining[\indexNode]$ is zero, the node changes its behavior. 
For instance, before entering the processing state, the node sets $\durationRemaining[\indexNode]$ to $\durationProcessing$~seconds. In the processing state, the node completes the computation tasks and waits until $\durationRemaining[\indexNode]$ becomes zero to go to the next state. We illustrate the states of the $\indexNode$th node  in  \figurename~\ref{fig:periodicRoles}. It is worth noting that the node can stay in the same state by extending $\durationRemaining[\indexNode]$, which is exploited in Section~\ref{subsubsec:jointSyncSlot}. Also, the state transitions based on $\durationRemaining[\indexNode]$ can be easily implemented in practice \cite{codeGitHubLoRaQuake}.

\begin{figure}[t]
	\centering
	\includegraphics[width = 3.0in]{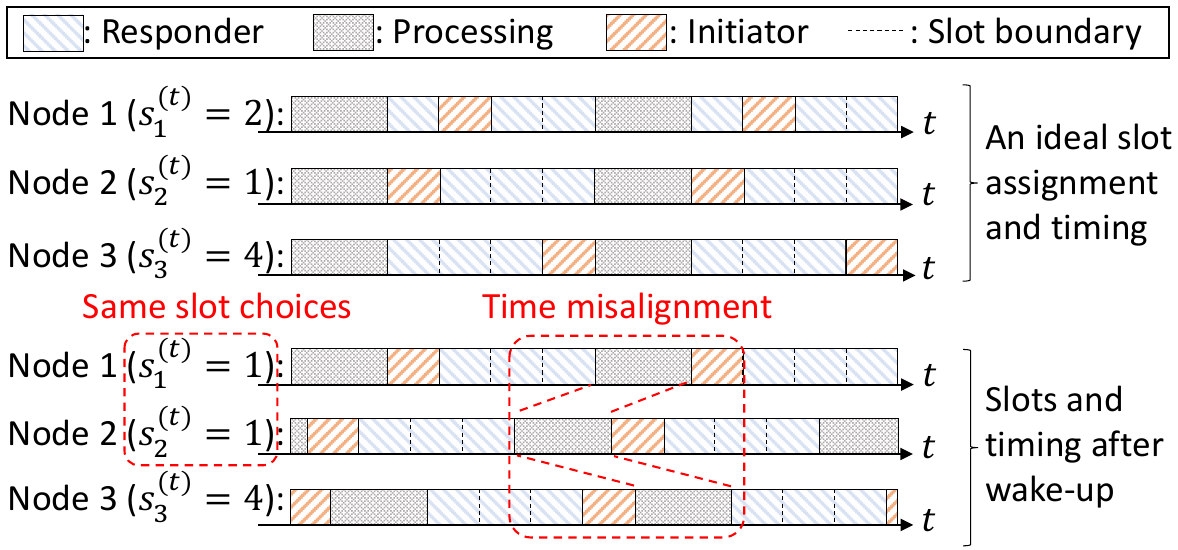}
	\caption{The nodes can choose the same slot index and they are out-of-sync since they wake up at different times  almost surely  ($\numberOfSlots=4$).}
	\label{fig:alignmentProblem}
\end{figure}
With \ac{PAR}, all nodes run the same cycle in their microprocessors as there is no hierarchical difference among the nodes. However, since the nodes wake up  at different instants almost surely, the processing and communication durations of the nodes are not aligned. Also, the slot indices  are chosen randomly initially. Hence, as exemplified in \figurename~\ref{fig:alignmentProblem} for $\numberOfSlots=4$, the collisions or missed packets can occur arbitrarily, and the nodes may not discover or communicate with their neighbors. 
 Suppose that global time synchronization is achieved and \ac{PAR} is used in the network. The slot assignment problem can then be expressed as
\begin{align}
	\{\nodeSlotOpt[\indexNode][\globalTime]|\forall\indexNode\}=\arg\min_{\{\nodeSlot[\indexNode][\globalTime]|\forall\indexNode\}} \sum_{\indexNode=1}^{\numberOfNodesTotal}\int_{\globalTime-\durationPacket}^{\globalTime}\indicatorFunction[{\functionConflict[{\indexNode}][{\globalTime'}]>1}]d\globalTime'~,
	\label{eq:optProblem}
\end{align}
for $\functionConflict[{\indexNode}][{\globalTime}] = \left|\{\nodeID[\indexNodeOther]|\nodeID[\indexNodeOther]\in\neighborSetR[\indexNode],
	\nodeState[\indexNodeOther]=\stateInit
	\}\right|
	$, where the function $\indicatorFunction[\cdot]$ results in $1$ if its argument holds, otherwise, it is $0$.
The function $\functionConflict[{\indexNode}][{\globalTime}]$ quantifies the number of transmitting nodes such that their signals reach the $\indexNode$th node at time $\globalTime$. If $\functionConflict[{\indexNode}][{\globalTime'}]>1$ for $\globalTime'\in[\globalTime-\durationPacket,\globalTime)$, the $\indexNode$th node can receive none of the transmitted packets on this duration for a synchronized network. The problem in \eqref{eq:optProblem} is combinatorial and it is not trivial to solve even for a coordinated static network. In this work, we address the same slot assignment problem in a distributed manner when there is no global time synchronization and the nodes do not know their neighbor sets.

\section{Proposed Methods}

\subsection{Joint time-and-slot adjustment and neighbor discovery}
\label{subsubsec:jointSyncSlot}


For a node to find its neighbors while addressing the synchronization and slot assignment problems, we propose a "listen-and-adjust" protocol relying on local adjustments that occur at any node that receives a beacon signal from one of its neighbors. Hence,  the proposed protocol inherently runs in parallel at all nodes. We introduce the following rules:
\begin{itemize}[    \setlength{\IEEElabelindent}{0in}
	]
\item If the $\indexNode$th node is an initiator, it transmits a beacon signal that indicates its \ac{ID} and slot index and its neighbors' \acp{ID} and slot indices, i.e., $\{(\nodeID[\indexNode], \nodeSlot[\indexNode][\globalTime]),\{(\nodeID[\indexNodeOther], \nodeSlot[\indexNodeOther][\globalTime])|\nodeID[\indexNodeOther]\in\neighborSetRX[\indexNode][\globalTime]\}\}\}$. We set $\durationPacket=\durationBeacon$ seconds, where $\durationBeacon$ is the duration of the beacon signal.

\item If the $\indexNode$th node is a responder and can decode the $\indexNodeOther$th node's beacon signal:
\begin{itemize} [   \setlength{\IEEElabelindent}{0in}
	]
 \item Adding nodes to neighbor lists: The $\indexNode$th node adds $\nodeID[\indexNodeOther]$ to the set $\neighborSetRX[\indexNode][\globalTime]$.
 If $\nodeID[\indexNode]$ is listed as a neighbor in the beacon, i.e., $\nodeID[\indexNode]\in\neighborSetRX[\indexNodeOther][\globalTime]$, (implying that the $\indexNodeOther$th node can decode the $\indexNode$th node's packets), $\nodeID[\indexNodeOther]$ is also added to $\neighborSetTX[\indexNode][\globalTime]$. 
 \item Removing the nodes from neighbor lists: To keep their neighbor lists up-to-date (i.e., tracking the changes in the graph), the $\indexNode$th node removes a node from their $\neighborSetRX[\indexNodeOther][\globalTime]$ and $\neighborSetTX[\indexNode][\globalTime]$, if its beacon signal is not heard for $\nForget(\durationProcessing+\durationComm)$~seconds.
 \item Slot adjustment: If  $\nodeSlot[\indexNode][\globalTime]$ is reported to be utilized at another node (i.e.,   $\nodeSlot[\indexNode][\globalTime]\in\{\nodeSlot[\indexNodeOther][\globalTime]\cup\{\nodeSlot[\indexNodeAnother][\globalTime]|\nodeID[\indexNodeAnother]\in\neighborSetRX[\indexNodeOther][\globalTime]\}\}$ -- slot index collision), the $\indexNode$th node chooses another slot from $\{1,\mydots,\numberOfSlots\}\backslash\{\nodeSlot[\indexNodeAnother][\globalTime]|\nodeID[\indexNodeAnother]\in\neighborSetRX[\indexNode][\globalTime]\cup\neighborSetRX[\indexNodeOther][\globalTime]\}$, randomly. If there is no available slot, collisions are inevitable, and  the slot index is kept the same.
 \item Local time synchronization: After the slot adjustment, the $\indexNode$th node re-calculates $\durationRemaining[\indexNode]$ as
\end{itemize}
\end{itemize}
\begin{small}
\begin{align}
	\small
	&\durationRemaining[\indexNode] =
	\begin{cases}
		(\nodeSlot[\indexNode][\globalTime]-\nodeSlot[\indexNodeOther][\globalTime]-1)\durationSlot+\durationDiff & \nodeState[\indexNode]=\stateResOne,\nodeSlot[\indexNode][\globalTime]>\nodeSlot[\indexNodeOther][\globalTime],  \\
		(\numberOfSlots+\nodeSlot[\indexNode][\globalTime]-\nodeSlot[\indexNodeOther][\globalTime]-1)\durationSlot&\\~~~~~~~~~~~~~~~~~+\durationProcessing+\durationDiff& \nodeState[\indexNode]=\stateResOne, \nodeSlot[\indexNode][\globalTime]<\nodeSlot[\indexNodeOther][\globalTime],\\
		(\numberOfSlots-\nodeSlot[\indexNodeOther][\globalTime])\durationSlot+\durationDiff & \nodeState[\indexNode]=\stateResTwo,
	\end{cases}.
	\label{eq:timer}
\end{align}
\end{small}


\begin{figure*}[t]
	\centering
	{\includegraphics[width=\figuresizeLarge]{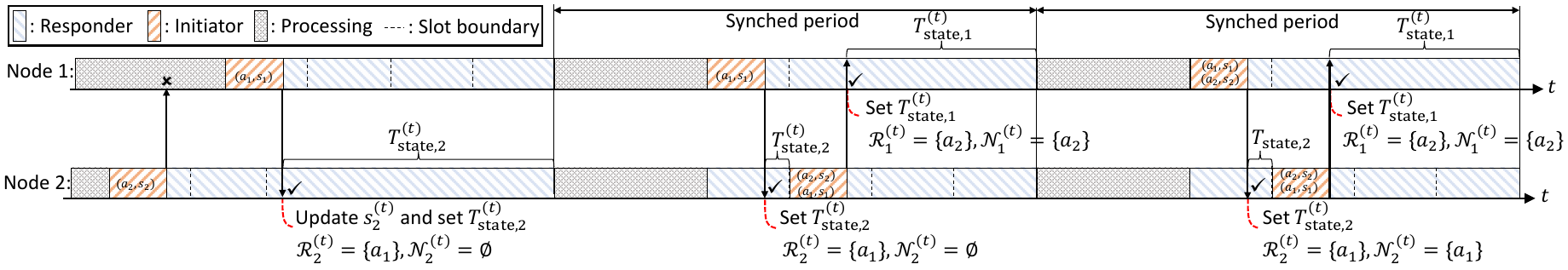}
	} 
	\caption{An example of the proposed time-and-slot adjustment and neighbor discovery protocol ($\numberOfNodesTotal=2$, $\numberOfSlots=4$). }
	\label{fig:proposedMethodExample}
\end{figure*}
In \figurename~\ref{fig:proposedMethodExample}, we provide an example illustrating how the proposed protocol works for $\numberOfNodesTotal=2$~nodes and $\numberOfSlots=4$~slots. Assume that the slot indices are $\nodeSlot[1][\globalTime]=1$ and $\nodeSlot[2][\globalTime]=1$ (i.e., the slot indices collide), $\neighborSetTX[1][\globalTime]$, $\neighborSetRX[1][\globalTime]$, $\neighborSetTX[2][\globalTime]$, and $\neighborSetRX[2][\globalTime]$ are empty sets, and the nodes are out-of-sync for $\globalTime=0$.  The first beacon (indicating $(\nodeID[2],\nodeSlot[2][\globalTime]=1)$) transmitted from Node 2 cannot be received by Node 1 as Node 1 is in the processing state. However, the first beacon (indicating $(\nodeID[1],\nodeSlot[1][\globalTime]=1)$) transmitted from Node 1 is received by Node 2. Hence, Node 2 adds $\nodeID[1]$ to $\neighborSetRX[2][\globalTime]$. Since the beacon does not report $\nodeID[2]$ as a neighbor of Node 1,  $\neighborSetTX[2][\globalTime]$ is still an empty set (i.e., Node 2 can listen Node 1, but it does not know if its signals can be decoded by Node 1). From the same beacon, Node 2 learns that Node 1 also uses the first slot index. Hence, it chooses another slot (i.e., $\nodeSlot[2][\globalTime]=2$) among three available slots to avoid collision. Subsequently, it re-calculates $\durationRemaining[2]$ as $3\durationSlot+\durationDiff$ seconds (the third case in \eqref{eq:timer}) and behaves as a responder before transitioning to the processing state. Hence, Node 2 becomes synchronous to the timing of Node 1. 
 Similarly, for the second beacon (indicating $(\nodeID[1],\nodeSlot[1][\globalTime]=1)$) transmitted   from Node 1, Node 2  re-adjusts $\durationRemaining[2]$  (the first case  in \eqref{eq:timer}) and waits for $\durationDiff$ seconds. The second beacon transmitted from Node 2 indicates  $(\nodeID[2],\nodeSlot[2][\globalTime]=2)$ and $(\nodeID[1],\nodeSlot[1][\globalTime]=1)$. Hence, Node 1 updates $\neighborSetRX[1][\globalTime]$ and also adds $\nodeID[2]$ to $\neighborSetTX[1][\globalTime]$ as its ID is listed in the beacon. It also calculates $\durationRemaining[1]$ (the third case  in \eqref{eq:timer}) as $2\durationSlot+\durationDiff$ seconds. Similarly, Node 2 adds $\nodeID[1]$ to $\neighborSetTX[2][\globalTime]$ after $\nodeID[2]$ is listed in the third beacon transmitted from Node 1. 
 In the following periods, the nodes re-calculate $\durationRemaining[1]$ and $\durationRemaining[2]$ when they receive a beacon signal, and they stay in sync with each other. 
 
Note that  if two nodes are in sync and choose the same slots, their signals can interfere with each other for all periods  for $\probabilityInitiator=1$, and the other nodes cannot learn and broadcast that the corresponding slot is utilized. For $\probabilityInitiator<1$, such collisions can be resolved and the slot information can be learned. Secondly,
the proposed protocol addresses the hidden node problem since  the nodes that receive a beacon packet avoid using the reported slots in the beacon.

\subsection{Transferring the sensing messages to the reference nodes}

We consider a similar approach to the Bellman-Ford routing~\cite{Bertsekas_1992}. Let us define the {\em hopping number} of the $\indexNode$th node, i.e., $\nodeHoppingNumber[\indexNode]$, as the number of hops to reach a reference node, calculated at time $\globalTime$. By  definition, the hopping number of a reference node is always $0$. For $\globalTime=0$, the sensing nodes do not know their hopping numbers and we set $\nodeHoppingNumber[\indexNode]$ to $\hoppingNumberMax$ if the $\indexNode$th node does not know its hopping number, where $\hoppingNumberMax$ is a pre-determined number larger than the maximum degree of separation of the graph. We extend the  rules discussed in Section~\ref{subsubsec:jointSyncSlot} as follows:
\begin{itemize}[   \setlength{\IEEElabelindent}{0in}
	]
	\item Broadcasting hopping number: If the $\indexNode$th node transmits a beacon signal, it indicates $\nodeHoppingNumber[\indexNode]$ in the beacon.
	\item Calculating hopping number:  If the $\indexNode$th node decodes the beacon signal transmitted from the $\indexNodeOther$th node, it registers $\nodeHoppingNumber[\indexNodeOther]$. It re-calculates $\nodeHoppingNumber[\indexNode]$ as
	\begin{align}
		\nodeHoppingNumber[\indexNode] =
		 1+ \min_{\substack{\nodeID[\indexNodeAnother]\in\neighborSetTX[\indexNode][\globalTime],\nodeHoppingNumber[\indexNodeAnother]<\hoppingNumberMax-1} } \nodeHoppingNumber[\indexNodeAnother]~,
	\end{align}
	if $\{\indexNodeAnother|\nodeID[\indexNodeAnother]\in\neighborSetTX[\indexNode][\globalTime],\nodeHoppingNumber[\indexNodeAnother]<\hoppingNumberMax-1\}$ is not an empty set. Otherwise, $\nodeHoppingNumber[\indexNode]$ is $\hoppingNumberMax$.
	\item Identifying the forwarding node: If the $\indexNode$th node receives a message,  it forwards it to the node with the smallest hopping number in $\neighborSetTX[\indexNode][\globalTime]$, i.e.,
	$
	\nodeID[\indexNodeAnother]' = \arg\min_{\nodeID[\indexNodeAnother]\in\neighborSetTX[\indexNode][\globalTime] } \nodeHoppingNumber[\indexNodeAnother]
	$.
	If there are multiple nodes with the smallest hopping number, the node forwards the sensing message to a randomly chosen node with the smallest hopping number. 
\end{itemize}

Note that the node can forward the message to the one that has the highest \ac{RSSI} or \ac{SNR} or all of the nodes with the smallest hopping number to improve reliability. Also, if the nodes register the pair of the ID of the original sender of the node and the ID of the forwarding node, the messages can be transferred from the reference node to the originating node via the same route in the reverse direction. 

\section{Numerical Results}
\label{sec:numericalResults}
\def\SNRref{\gamma_{0}}
\def\dref{d_{0}}
\def\SNRmin{\gamma_{\rm m}}
\def\pathLossExponent{\eta}
\def\shadowingVariance{\sigma_^2}
\def\shadowingCoef[#1]{\phi_{#1}}
\def\smallScale[#1]{h_{#1}}
\def\powerImbalance[#1]{\psi_{#1}}
In this section, we assess the proposed methods with both simulations and a proof-of-concept demonstration. For simulations, we consider $\numberOfNodesReference=5$~nodes separated apart by 30~m and 25~m on the x-axis and y-axis, respectively. We assume that $\numberOfNodesInside=25$ sensing nodes are randomly distributed in an area where its size is 125~m by 100~m unless otherwise stated. For the large-scale fading, we assume that the path loss exponent is $\pathLossExponent=3.7$ and the variance of the log-normal shadowing coefficient between the $\indexNode$th and $\indexNodeOther$th node, i.e., $\shadowingCoef[\indexNode,\indexNodeOther]$ is 6 dB for $\shadowingCoef[\indexNode,\indexNodeOther]=\shadowingCoef[\indexNodeOther,\indexNode]$. For the small-scale fading, we consider Rayleigh fading, i.e., $\smallScale[\indexNode,\indexNodeOther]\sim\complexGaussian[0][1]$ for $|\smallScale[\indexNode,\indexNodeOther]|^2=|\smallScale[\indexNodeOther,\indexNode]|^2$. We assume that the minimum \ac{SNR} is $\SNRmin=-5$~dB to decode a packet  and the reference \ac{SNR} is $\SNRref=20$~dB if the distance between two nodes is $\dref=10$~meters. We model the transmit power imbalance of the $\indexNode$th node, i.e., $\powerImbalance[\indexNode]$, as a zero-mean Gaussian distribution with a variance of 3 dB. Based on this model, we obtain $\accessible[\indexNode][\indexNodeOther]$ as
\begin{align}
\accessible[\indexNode][\indexNodeOther]=
\begin{cases}
	1 &  \SNRref+\powerImbalance[\indexNodeOther]+\shadowingCoef[\indexNode,\indexNodeOther]+10\log_{10}\frac{\distance[\indexNode][\indexNodeOther]^{\pathLossExponent}|\smallScale[\indexNode,\indexNodeOther]|^2}{\dref^{\pathLossExponent}}>\SNRmin,\\
	0  &\text{otherwise},
\end{cases}~.
\nonumber
\end{align}
We choose $\durationBeacon=5$~ms, $\durationSlot=10$~ms, $\durationProcessing=10$~ms, and $\probabilityInitiator=\{0.25,0.5\}$, $\hoppingNumberMax=30$, and $\nForget=10$. The nodes awake up at random times between 0 and 100~ms. The behaviors of the nodes are modeled in MATLAB environment.

\begin{figure}
	\centering
	\subfloat[Number of victims over time for a regular deployment ($\numberOfSlots=12$).]{~~\includegraphics[width = 3.0in]{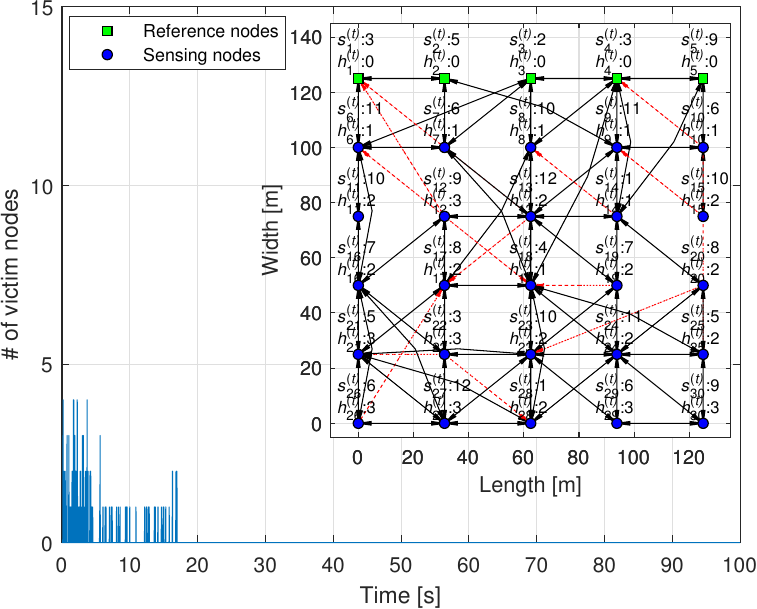}
	\label{subfig:regularDep}~~}
		\\
	\subfloat[Number of victims over time for a  random deployment ($\numberOfSlots=16$).]{~~\includegraphics[width =3.0in]{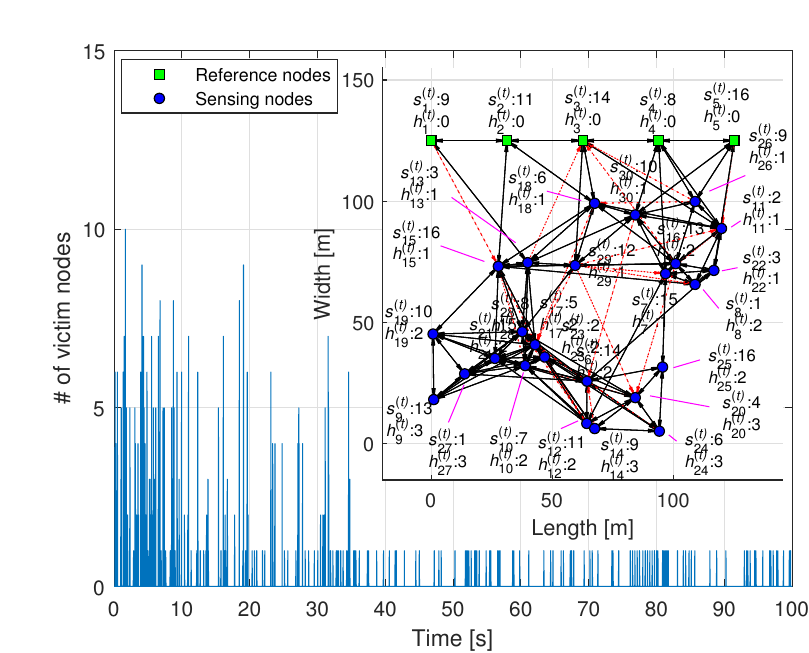}
		\label{subfig:randomDep}~~}
	\caption{The proposed  protocol reduces the collision events while resolving the hidden-node problems ($\probabilityInitiator=0.5$, a dashed arrow: a directional link,  a solid double-ended arrow: a bi-directional link).}	
	\label{fig:proposedMethodGraphExample}	
\end{figure}

In \figurename~\ref{fig:proposedMethodGraphExample}, we demonstrate the performance of the proposed protocol by evaluating the number of victim nodes (i.e., a node interfered by a neighboring node) over time, i.e., $\sum_{\indexNode=1}^{\numberOfNodesTotal}\indicatorFunction[{\functionConflict[{\indexNode}][{\globalTime}]>1}]$ in \eqref{eq:optProblem}, for two deployment scenarios. In \figurename~\ref{fig:proposedMethodGraphExample}\subref{subfig:regularDep}, we consider a rectangular tessellation. We also indicate the links between the nodes and mark the slots  and the hopping numbers chosen by the nodes at $\globalTime=100$~seconds for $\numberOfSlots=12$. For this configuration, the nodes reach a consensus at $\globalTime=17$~seconds and the number of victim nodes quickly reduces to $0$.  In \figurename~\ref{fig:proposedMethodGraphExample}\subref{subfig:randomDep}, we consider random deployment  for $\numberOfSlots=16$. Initially, we observe a large number of victim nodes. However, the protocol effectively mitigates the collision over time and there is only 1 victim node after $\globalTime=35$~seconds in this challenging graph. Note that
for both scenarios, the protocol  addresses the hidden node problems. For example, slot~$11$ is used by the $(6,9,24)$th nodes and none of their first and second degrees of neighbors use slot $11$ after the resolution in \figurename~\ref{fig:proposedMethodGraphExample}\subref{subfig:regularDep}. In \figurename~\ref{fig:proposedMethodGraphExample}\subref{subfig:regularDep} and \figurename~\ref{fig:proposedMethodGraphExample}\subref{subfig:randomDep}, we also observe that each node obtains its hopping number based on its shortest hopping path to one of the reference nodes. For example, in \figurename~\ref{fig:proposedMethodGraphExample}\subref{subfig:regularDep}, $\nodeHoppingNumber[12]=3$ as the minimum hopping number in $\neighborSetTX[12][\globalTime]=\{\nodeID[13],\nodeID[16],\nodeID[17]\}$ is $2$.

\begin{figure}[t]
	\centering
	{\includegraphics[width=3.0in]{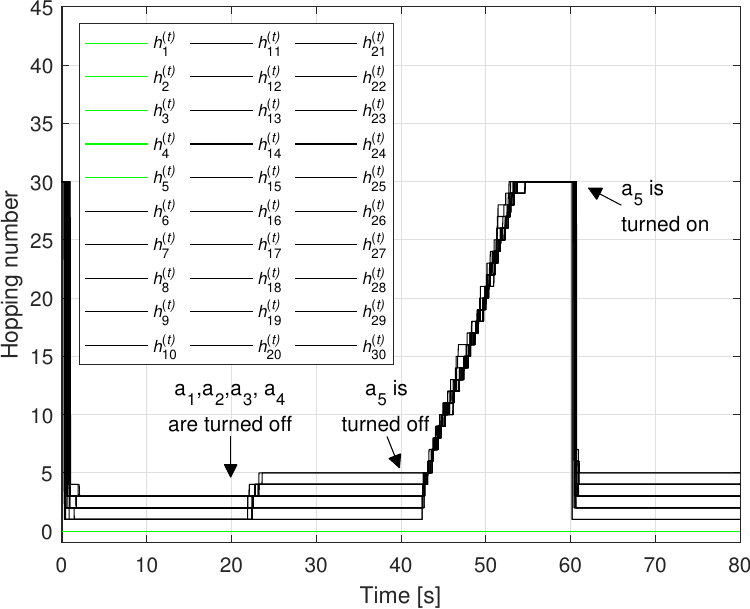}
	} 
	\caption{In this example, the reference nodes are turned off and turned on and all nodes adapt themselves by re-calculating their hopping numbers (green: reference nodes, black: sensing nodes).}
	\label{fig:healingMethodExample}
\end{figure}
\begin{figure}[t]
	\centering
	\subfloat[$\probabilityInitiator=0.5$.]{\includegraphics[width =3.0in]{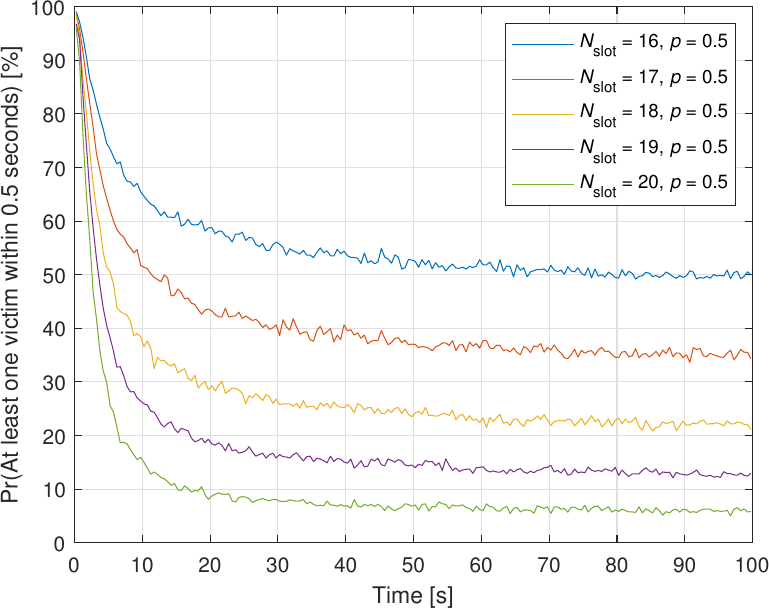}
		\label{subfig:figure_probCol_scenario3_p50}}\\
	\subfloat[$\probabilityInitiator=0.25$.]{\includegraphics[width =3.0in]{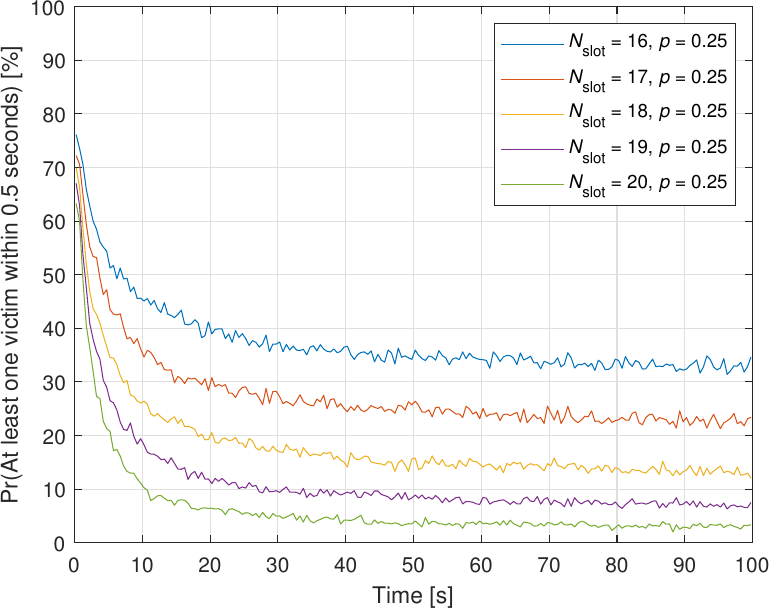}
		\label{subfig:figure_probCol_scenario3_p25}}\\
	\caption{The collision probability  reduces over time with proposed protocol.}
	\label{fig:probCol}
\end{figure}
In \figurename~\ref{fig:healingMethodExample}, we demonstrate the self-healing behavior of the network. We consider the same scenario in \figurename~\ref{fig:proposedMethodGraphExample}\subref{subfig:regularDep} and plot how the hopping numbers change over time. We turn off the first four reference nodes at $\globalTime=20$~seconds. At $\globalTime=22$~seconds, the nodes remove the turned-off nodes from their neighbor lists as they do not get any beacon from the turned-off nodes. Consequently, the sensing nodes re-adjust their hopping numbers within 2 seconds. At $\globalTime=40$~seconds, we also turn off the last reference node. Hence, there is no reference node in the network. As a result, all the nodes gradually increase their hopping numbers till they reach $\hoppingNumberMax=30$. We then turn on the fifth node and the nodes quickly discover the reference node and learn their hopping numbers.

In \figurename~\ref{fig:probCol}, we analyze  the probability of having at least one victim node within $[\globalTime-\aDuration/2,\globalTime+\aDuration/2)$ duration for a given $\globalTime$, i.e.,
$\probability[{\indicatorFunction[{(\sum_{\indexNode=1}^{\numberOfNodesTotal}\int_{\globalTime-\aDuration/2}^{\globalTime+\aDuration/2}\indicatorFunction[{\functionConflict[{\indexNode}][{\globalTime'}]>1}])d\globalTime'>0}]}=1]$ for the same scenario in \figurename~\ref{fig:proposedMethodGraphExample}\subref{subfig:randomDep} and set $\aDuration$ to $0.5$~seconds.  We consider 2000 realizations and run the experiments for $100$~seconds.  As can be seen from \figurename~\ref{fig:probCol}, the proposed protocol reduces the number of victim nodes over time. Also, as expected, increasing the number of slots reduces the collision events. In \figurename~\ref{fig:probCol}\subref{subfig:figure_probCol_scenario3_p50}, we use $\probabilityInitiator=0.5$. The collision probability is reduced from the range of 90\%-100\% to 10\%  within 20~seconds for $\numberOfSlots=20$. In \figurename~\ref{fig:probCol}\subref{subfig:figure_probCol_scenario3_p25}, we reduce $\probabilityInitiator$ to $0.25$ and the collision probability is lowered as compared to the first case. In this case, we observe that the probability is reduced from approximately 70\% to 10\%  within $10$ and $30$ seconds for $20$ and $19$ slots, respectively.

\begin{figure}[t]
	\centering
	\subfloat[The setup for the proof-of-concept demonstration.]{\includegraphics[width =2.6in]{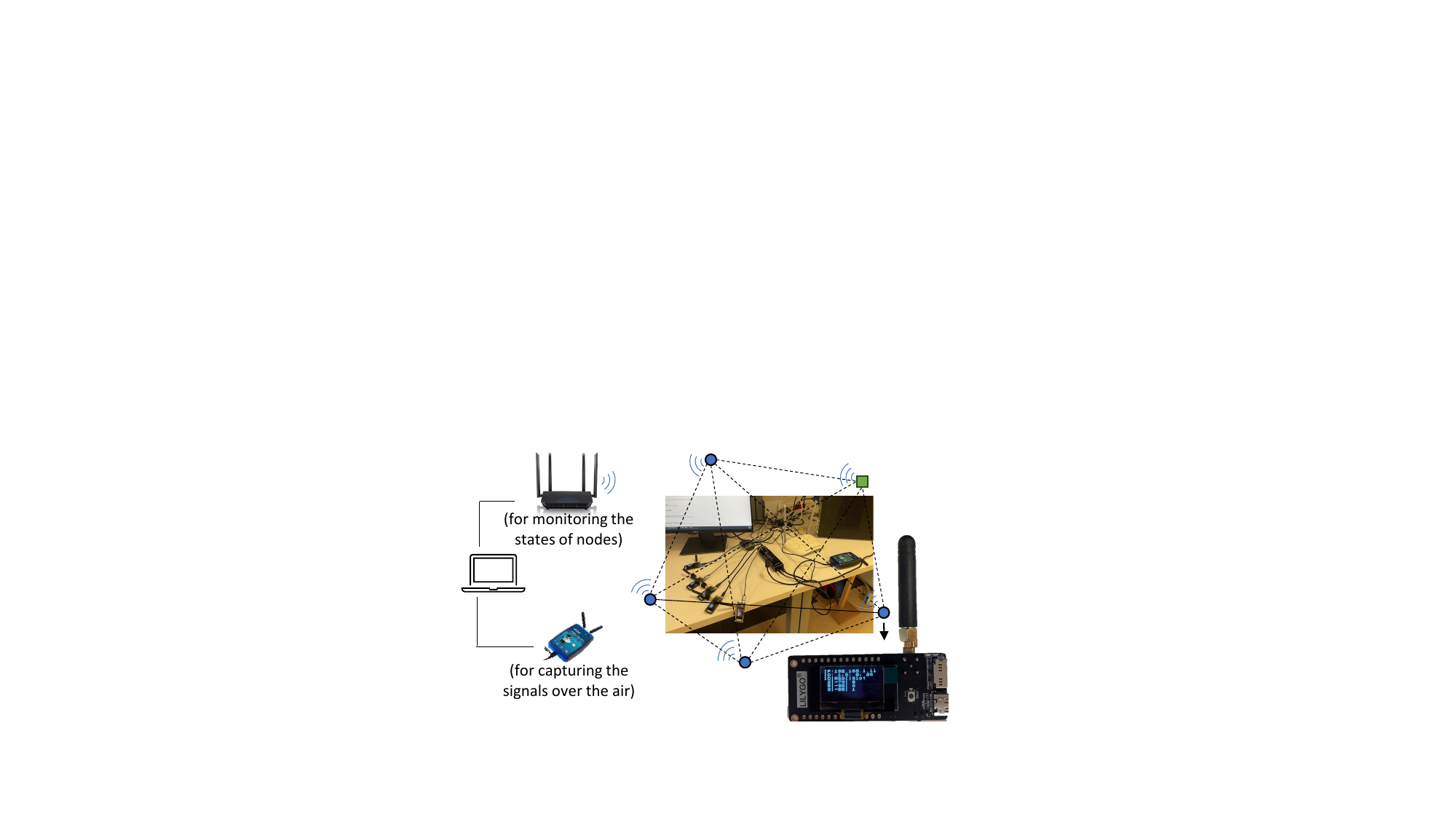}
	\label{subfig:setup}}\\	
	\subfloat[MATLAB GUI. The RSSI is shown by blue bars.  
	]{\includegraphics[width =2.9in]{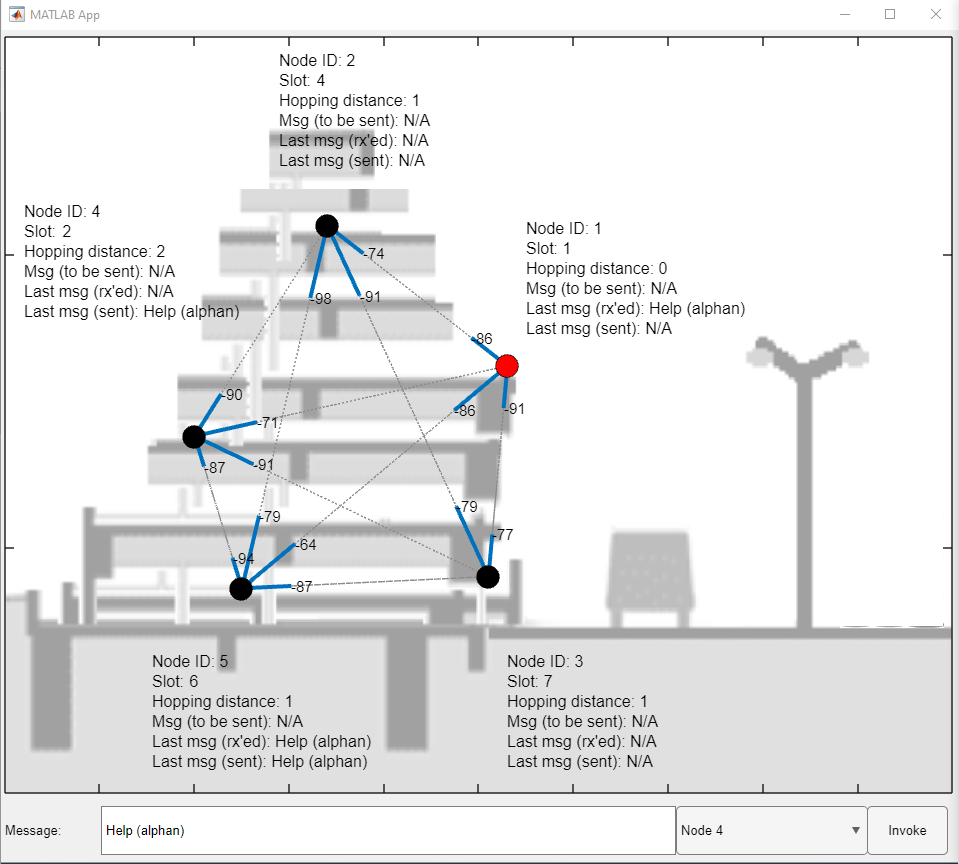}
		\label{subfig:gui}}\\
	\subfloat[An instant of the captured LoRa signals with GNU Radio. ]{\includegraphics[width =3in]{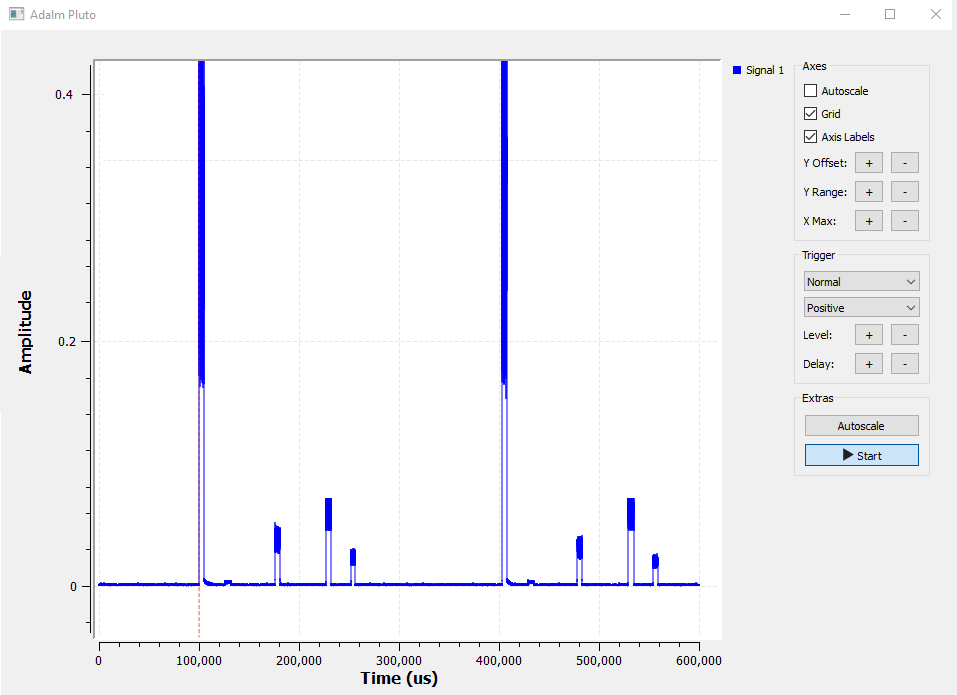}
	\label{subfig:adalm}}\\		
	\caption{Implementation ($\numberOfNodesInside=4$ and $\numberOfNodesReference=1$). The message sent from Node 4 reaches Node 1 over Node 5. The synchronization is achieved and the slots are chosen as $\nodeSlot[1][\globalTime]=1$, $\nodeSlot[2][\globalTime]=4$, $\nodeSlot[3][\globalTime]=7$, $\nodeSlot[4][\globalTime]=2$, and $\nodeSlot[5][\globalTime]=6$.}	
	\label{fig:implementation}
	\vspace{-3mm}
\end{figure}
We also implement the proposed methods by using 5 LILYGO \ac{LoRa} development boards using ESP32-S3 and Semtech SX1280 chipset as can be seen in \figurename~\ref{fig:implementation}  for $\numberOfNodesInside=4$ and $\numberOfNodesReference=1$ \cite{codeGitHubLoRaQuake}. In this setup, we acquire the status of each node over a Wi-Fi network during the processing duration and monitor the activity in the spectrum with an Adalm-Pluto via GNU Radio, as shown in \figurename~\ref{fig:implementation}\subref{subfig:setup}. We track the state of each node, its slot, ID, hopping number, and \ac{RSSI} of their neighbors over a \ac{GUI} in MATLAB  as shown in \figurename~\ref{fig:implementation}\subref{subfig:gui}. In the \ac{GUI}, the bars indicate the \ac{RSSI} in dBm. To create asymmetrical links, we intentionally add a $30$~dB attenuator between the antenna and the RF port  for Node~4 and Node~5. Hence, they can decode the packets, but their signals cannot be decoded by all nodes. With this setup, we demonstrate the slot assignments, time adjustments, and multi-hop routing. For the implementation, we set  $\durationSlot=25$~ms, $\durationProcessing=100$~ms, $\numberOfSlots=8$, $\nForget=50$, and $\hoppingNumberMax=127$. For the \ac{LoRa} packets, we use a spreading factor of $10$, a coding rate of $4/5$, and $812.5$~kHz bandwidth, respectively. As can be seen from \figurename~\ref{fig:implementation}\subref{subfig:gui} and \figurename~\ref{fig:implementation}\subref{subfig:adalm}, the nodes choose their slots as $\nodeSlot[1][\globalTime]=1$, $\nodeSlot[2][\globalTime]=4$, $\nodeSlot[3][\globalTime]=7$, $\nodeSlot[4][\globalTime]=2$, and $\nodeSlot[5][\globalTime]=6$ and synchronization in the entire network is achieved without any \ac{GPS} signal and it is well-aligned in the time domain. In the demo, the hopping number of Node~4 is $2$ because the minimum hopping number among the nodes in $\neighborSetTX[5][\globalTime]$ (i.e., Node 5) is $1$. Hence, if Node~4 is triggered to transmit a message, the corresponding message is forwarded to Node~5 first. Afterward, the Node 5 forwards the message to the reference node, i.e., Node~1.

\section{Concluding Remarks}
In this study, we present a self-healing mesh network that does not rely on global-time synchronization. In the network, all nodes adjust their slots and timings via the received beacon signals and synchronize themselves with their neighbors locally. Also, the nodes obtain the forwarding nodes on the optimal routes without knowing the communication graph. Through comprehensive simulations, we show that the proposed protocol effectively reduces collision events, resolves hidden node problems, and tracks the changes in the network. We also implement the proposed protocols by using LoRa SX1280 chipsets and provide proof-of-concept results. 
In particular, the proposed protocol can be useful for building a low-cost mesh network in challenging scenarios where the \ac{GPS} signals cannot penetrate.
Future work will analyze the data rate and energy efficiency of the proposed scheme while incorporating the sensing features of nodes.

\bibliographystyle{IEEEtran}
\bibliography{references}

\end{document}